# First-Principles Study of Lattice Thermal Conductivity of Td-WTe$_2$

Gang Liu[1*], Hong Yi Sun[1*], Jian Zhou[2+], Qing Fang Li[1,3+], X. G. Wan[1]

[*] These authors contributed equally to this work.

[1] National Laboratory of Solid State Microstructures, College of Physics, Nanjing University, Nanjing, 210093, China

[2] National Laboratory of Solid State Microstructures and Department of Materials Science and Engineering, Nanjing University, Nanjing 210093, China

[3] Department of Physics, Nanjing University of Information Science & Technology, Nanjing 210044, China

The structural and thermal properties of bulk Td-WTe$_2$ have been studied by using first-principles calculations. We find that the lattice thermal conductivity of WTe$_2$ is anisotropic, with the highest value of 9.03 Wm$^{-1}$K$^{-1}$ along a-axis and lowest one of 0.46 Wm$^{-1}$K$^{-1}$ along the c-axis at 300 K. Our calculated size dependent thermal conductivity shows that nanostructuring of WTe$_2$ can possibly further decrease the lattice thermal conductivity. Such extremely low thermal conductivity, even much lower than WSe$_2$, makes WTe$_2$ attractive for use as thermal-insulation material for thermoelectric devices.

[+] Correspondence and requests for materials should be addressed to J. Zhou (e-mail: zhoujian@nju.edu.cn); or to Q.F. Li (e-mail: qingfangli@nuist.edu.cn)

# Introduction

Tungsten telluride (Td-WTe$_2$), as a typical transition-metal dichalcogenide (TMD), has attracted great attention recently owing to a number of interesting physical properties, [1-10] such as non-saturating giant positive magnetoresistance [1], superconductivity [7,8] and high carrier mobilities. [4] Unfortunately, the extremely large positive magnetoresistance can only be observed at low temperature which limits the applications of WTe$_2$ magnetorestance devices. The thermal transport properties are a key consideration to accelerate the application of WTe$_2$, which are likely to demonstrate thermally limited performance. In addition, thermal conductivity plays a crucial role in defining thermoelectric efficiency, thus the thermal transport of materials has become an important subject to design the high-efficiency thermoelectric materials which can convert waste heat into usable electrical energy.

Recently, the thermal properties of TMD materials have attracted great interest due to their potential applications in ultralow thermal conductivity devices [11] and thermoelectric devices. [12-16] The performance of thermoelectric materials depends on the figure of merit ZT. [17] One effective way to increase ZT is to reduce the thermal conductivity without affecting electronic conductivity. [18] Moreover, ultralow thermal conductivity is required to prevent the back-flow of heat from the hot end to the cool one. Therefore, it is crucial to search for ultralow thermal conductivity materials in thermoelectric applications. It was recently found that the cross-plane thermal conductivity ($\kappa$) of disordered WSe$_2$ thin films is as low as 0.05 Wm$^{-1}$K$^{-1}$ [11] which is the lowest $\kappa$ ever reported for a dense solid to this date. According to the classical theory, [19, 20] WTe$_2$ should possess lower thermal conductivity than WSe$_2$ due to the heavier atom mass and lower Debye temperature in WTe$_2$. However, unlike the electronic properties of WTe$_2$, which have been intensively explored, the study on the thermal properties is still in its infancy. [21-24] The experiment measurement showed the thermal conductivity of polycrystalline WTe$_2$ is 0.8 Wm$^{-1}$K$^{-1}$, [24] which didn't exhibit the anisotropy of thermal transport. Besides, the polycrystalline samples must have many boundary scattering which can not give the intrinsic thermal conductivity of WTe$_2$.

In this paper, we have calculated the isochoric specific heat, Debye temperature, and the intrinsic lattice thermal conductivity from the first principles calculations. We find that the thermal conductivity of WTe$_2$ is anisotropic, and much lower than that of WSe$_2$. The size-dependent thermal conductivity of WTe$_2$ is also discussed.

## Computational details

The first-principles calculations are performed by using the density functional theory (DFT) as implemented in the Vienna ab-initio simulation package (VASP). [25-27] The Perdew-Burke-Ernzerhof (PBE) of generalized gradient approximation (GGA) is chosen as the exchange-correlation functional. [28] To take into account van der Waals forces (vdW), which are expected to play a crucial role in bulk WTe$_2$, the empirical correction scheme of Grimme (DFT+D2) [29,30] are used in our calculation. The cutoff energy of planewaves is set as 600 eV, and a Monkhorst-Pack [31] k-mesh of 14 × 7 × 3 is used to sample the Brillouin zone in the structure optimization. The internal coordinates and lattice constants are optimized until the atomic forces became less than $10^{-3}$ eV/Å.

In the calculation of phonon dispersion, a 3 × 2 × 1 supercell containing 72 atoms is constructed to ensure the convergence and a 5 × 4 × 3 k-mesh for the Brillouin zone sampling is used. The phonon dispersion and density of states (DOS) is obtained from the finite displacement method as implemented in the PHONOPY package. [32] An 84 × 54 × 18 q-mesh is used in the specific heat capacity and thermal conductivity calculations.

Based on the phonon dispersion, the lattice thermal conductivity $\kappa$ of a crystal at finite temperature $T$ can be calculated from the following formulas:

$$\kappa = \frac{1}{N_0 \Omega} \sum_{\lambda, q} [\mathbf{v}_\lambda(\mathbf{q}) \cdot \mathbf{t}]^2 \tau_\lambda(\mathbf{q}) C_{ph}(\omega_\lambda), \qquad (1)$$

$$C_{ph}(\omega) = \frac{(\hbar\omega)^2}{k_B T^2} \frac{\exp(\hbar\omega/k_B T)}{[\exp(\hbar\omega/k_B T) - 1]^2}, \qquad (2)$$

$$\frac{1}{\tau_\lambda(\mathbf{q})} = 2\gamma_\lambda^2(\mathbf{q}) \frac{k_B T}{Mv^2} \frac{\omega_\lambda^2(\mathbf{q})}{\omega_m}, \qquad (3)$$

where $\Omega$ is the volume of the unit cell, $N_0$ is the total number of q points in the first Brillouin zone, $\lambda$ is the index of the phonon modes and runs over all the phonon branches, t is a unit vector in the direction of thermal gradient $\nabla T$, $\mathbf{v}_\lambda(\mathbf{q})$ is the group velocity, and $C_{ph}(\omega_\lambda)$ is the contribution of phonon modes to the specific heat. [33-35] $\tau_\lambda(\mathbf{q})$ is the intrinsic phonon relaxation time associated with the three-phonon umklapp scattering, where $M$ is the averaged atomic mass, $\omega_m$ is the Debye frequency, $v$ is the averaged sound velocity around the Brillouin zone center, $\gamma_\lambda(\mathbf{q})$ is Grüneissen parameter defined as $\gamma_\lambda(\mathbf{q}) = -\frac{V}{\omega_\lambda(\mathbf{q})} \frac{\partial \omega_\lambda(\mathbf{q})}{\partial V}$, and $V$ is the equilibrium volume. This relax time expression was derived by Klemens based on the time-dependent perturbation theory. [33,36,37] The interactions of phonon and electron can be neglected safely as the electronic density of states near the Fermi surface is quite small. [7]

The cumulative thermal conductivity is calculated in order to examine the size dependence of $\kappa$. When calculating the thermal conductivity, we can only sum the modes whose phonon mean-free paths (MFP) are shorter than a threshold by rewriting the equation (1): [38]

$$\kappa^{direc}(L) = \frac{1}{N_0 \Omega} \sum_{\lambda,\mathbf{q}}^{l_\lambda^{direc}(\mathbf{q}) \leq L} \mathbf{v}_\lambda^{direc}(\mathbf{q}) l_\lambda^{direc}(\mathbf{q}) C_{ph}(\omega_\lambda), \quad (4)$$

$$l_\lambda^{direc}(\mathbf{q}) = \mathbf{v}_\lambda^{direc}(\mathbf{q}) \tau_\lambda(\mathbf{q}), \quad (5)$$

where $l_\lambda^{direc}(\mathbf{q})$ is the MFP for each mode in a direction, and L is a certain MFP threshold, and $\kappa^{direc}(L)$ is so-called the cumulative thermal conductivity. We can study the relationship between cumulative thermal conductivity and the MFP threshold L to estimate the size-dependent thermal conductivity, which can give more information for experiments.

## Results and discussion

The space group of bulk WTe$_2$ is Pnm2$_1$. The octahedron of Te is slightly distorted and the W atoms displaced from their ideal octahedral sites, forming zigzag metal-metal chains along a-axis [39] as shown in Fig. 1. The crystal structure is fully optimized using the GGA and GGA+vdW methods. The calculated structural parameters are listed in Table 1. Taking the experimental ambient volume [39] as a reference, we find the difference between calculated and experimental volumes of 15.83% in GGA method and 2.89% in GGA+vdW method. Obviously, the GGA+vdW scheme performs much better. From Table I, we can find that the inclusion of vdW interaction reduces all the lattice parameters (a, b, and c). But the largest difference between the two methods is found for the c parameter, i. e., the direction perpendicular to the WTe$_2$ layers. The distance between two adjacent layers (d in Fig. 1) is also better described by GGA+vdW. Therefore, we conclude that the vdW dominate the interlayer interaction and is essential in order to obtain the accurate lattice parameters in WTe$_2$. Therefore all of the following results are based on the GGA+vdW method.

The calculated phonon dispersion and DOS are shown in Fig.2. The phonon frequencies are in the range of 0~232 cm$^{-1}$. It can be found that all of the phonon frequencies are positive in the Brillouin zone, which indicates that the structure of WTe$_2$ is dynamically stable. Since the primitive cell of WTe$_2$ contains twelve atoms, thirty-six independent vibration modes can be found, in which three are acoustic modes (two transverse and one longitudinal) and the remaining modes correspond to optical ones. The phonon dispersions are well consistent with previous reports. [7]

According to the results of the phonon dispersion, the isochoric specific heat capacity $C_v$ of bulk WTe$_2$ is calculated. Within the harmonic approximation, $C_v$ can be calculated using the formulas [40] $C_v = \sum_{\lambda,\mathbf{q}} C_{ph}(\omega_\lambda(\mathbf{q}))$. The temperature dependent heat capacity is presented in Fig. 3(a). It can be seen the isochoric specific heat capacity shows the expected $T^3$ law behavior in the low–temperature limit, and attain the saturation value (74.8 Jmol$^{-1}$K$^{-1}$), which is known as Dulong-Petit classical limit. Compared with the experimental constant-pressure specific heats $C_p$ [22] as shown in Fig. 3, it is found that the $C_p$ are systemically larger than the $C_v$ which is related to the thermal expansion caused by anharmonicity effects [41].

The Debye temperature $\theta_D$ is known as an important fundamental parameter closely related to many physical properties such as specific heat, elastic constant, and melting point. [42] We obtain the Debye temperature based on Debye model by fitting the calculated specific heat capacities. Debye temperature $\theta_D$ as a function of temperature is exhibited in Fig. 3(b). As the temperature varies from 0 to 500 K, $\theta_D$ increases quickly and then reaches a constant about 263 K. The zero temperature $\theta_D(T \to 0)$ is 137.0 K, which is in good agreement with the experimental one (133.8 K). [22]

The calculated thermal conductivities of $WTe_2$ and $WSe_2$ at 300 K are listed in Table 2 for comparison. For $WSe_2$, our results agree reasonably with the recent DFT calculation [16] and well consistent with the experimental results [43]. As we expected, the thermal conductivities of $WTe_2$ is lower than that of $WSe_2$. For instance, the lowest $\kappa$ of $WTe_2$ is 0.46 $Wm^{-1}K^{-1}$ along [001] direction, while the lowest value for $WSe_2$ is about two times higher at 0.90 $Wm^{-1} K^{-1}$ at the same direction. The thermal conductivity of $WTe_2$ is much lower than that of $WSe_2$ because of the heavier atom mass and lower Debye temperature in $WTe_2$. We estimate the averaged thermal conductivity is about 1.35 $Wm^{-1}K^{-1}$ using the Matheissen's rule. The calculated result is slightly larger than the experimental value (0.8 $Wm^{-1}K^{-1}$) [24], which is probably due to the defect and/or boundary scattering effects in polycrystalline samples.

In order to compare the temperature dependent $\kappa$ of $WTe_2$ and $WSe_2$, we present the calculated lattice thermal conductivities of $WTe_2$ and $WSe_2$ from 200 to 500 K in Fig. 4. In our calculation, only the intrinsic phonon-phonon scattering process based on phonon–phonon umklapp scattering mechanism [33, 36, 37] is considered. The extrinsic scattering from the defects and boundaries are not included, which should be strongly depending on microstructures of materials. However, we will use the cumulative thermal conductivity by different MFP threshold to estimate the size effect in real materials later. From Fig. 4, we can see that the thermal conductivity of $WTe_2$ is much lower than that of $WSe_2$ not only at the room temperature, but also in the whole calculated temperature range. It is also shown that both of their lattice thermal conductivities decrease with the rise of temperature following a $T^{-1}$ relationship due to the stronger phonon-phonon scattering at the higher temperature.

Besides, we find the obvious anisotropy of thermal conductivity in both materials. The thermal conductivities in [001] direction are much lower than those propagating on (001) plane. This kind of anisotropy of thermal conductivity between in-plane and cross-plane is due to their vdW bound layered structure which has also been found in other layered materials. [44] The in-plane thermal conductivity in WTe$_2$ is a factor of ~19 larger than the c-axis thermal conductivity, which is slightly higher than that of WSe$_2$. The in-plane thermal conductivities of WTe$_2$ are also a little anisotropic while WSe$_2$ shows an isotropous which is ascribed to their different crystal symmetry. Fig. 5 shows the in-plane thermal conductivity $\kappa$ of WTe$_2$ at 300 K to understand the anisotropy of in-plane thermal conductivity. It can be seen the in-plane thermal conductivity is 7.69 ~ 9.03 Wm$^{-1}$K$^{-1}$ at 300 K, and the maximum is along [100] direction while the minimum is along [010] direction.

To further understand the anisotropy of thermal conductivity, we have calculated the group velocities of phonons of WTe$_2$, which are closely relevant to the lattice thermal conductivity of a material. The group velocities along [100], [010] and [001] directions, as a function of frequency, are plotted in Fig. 6. It can be seen that the group velocities decreases with the increase of phonon frequency along each direction, which is because the optical modes are usually less dispersive than the acoustic ones. In addition, the phonon group velocities along [001] direction are significantly lower than the other two directions because of the weak inter-layer vdW interactions along the [001] direction. Therefore, the phonon velocity is anisotropic which has significantly effect on the lattice thermal conductivity. [45] The larger phonon velocities in the direction [100] and [010] result in a larger lattice thermal conductivity in these directions, and smaller one in [001] is responsible for smaller thermal conductivity.

An important question on the thermal conductivity of real materials is the size dependence. In order to study the size dependent thermal conductivity, we calculated the cumulative thermal conductivities of WTe$_2$ along [100], [010] and [001] directions at 300 K as shown in Fig. 7. The total accumulation for each direction fistly increases as L (MFP threshold), then gradually approaches plateau. It can be seen that the cumulative thermal conductivities in direction [100] and [010] are more size-dependent than that in direction [001]. The cumulative thermal conductivity in [100] direction decrease quickly when the L is less than 100 nm, while the critical point of such quick descrease is about 75 and 25 nm for [010] and [001], respectively. Therefore, the nanostructuring approach is a promising way to reduce the lattice thermal conductivity. However, the electrical conductivity may decreases with the nanostructuring which decreases the ZT, it can keep electronic conductivity by doping such as Mo to obtain a high thermoelectric efficiency.

## Conclusion

We have investigated the structural properties and the intrinsic lattice thermal conductivity of bulk WTe$_2$ from first principles. It is found that vdW play an important role in the inter-layer interaction of WTe$_2$. Our calculations indicate that the thermal conductivity of WTe$_2$ is anisotropic, and the highest thermal conductivity (9.03 W m$^{-1}$ K$^{-1}$ at 300 K) is along a-axis, while the lowest one is along c-axis (0.46 W m$^{-1}$ K$^{-1}$ at 300 K). The anisotropic group velocities of phonon are responsible for the anisotropy of thermal conductivity. In addition, the WTe$_2$ has the ultralow low cross-plane thermal conductivity which is even lower than that of WSe$_2$. We also studied the size dependent thermal conductivity of WTe$_2$. Our results indicate that introducing nanostructure can possibly further decrease the lattice thermal conductivity by one order of magnitude in WTe$_2$. Our work shield more light on potential applications of WTe$_2$ in thermal-insulation material for thermoelectric devices.

## Acknowledgements

This work was supported by the National Key Project for Basic Research of China (Grant nos. 2015CB659400), NSFC (Grant nos. 11474150) and China Postdoctoral Science Foundation (2014M551544).

Table 1. Calculated structural parameters of bulk WTe$_2$. Experimental ones are listed for comparison. *d* is the nearest interlayer distance shown in Fig. 1.

| Lattice parameter(Å) | GGA | GGA+vdW | Expt.[a] |
|---|---|---|---|
| *a* | 3.502 | 3.500 | 3.477 |
| *b* | 6.332 | 6.291 | 6.249 |
| *c* | 15.910 | 14.232 | 14.018 |
| *d* | 3.764 | 2.906 | 2.868 |
| V (Å$^3$) | 352.82 | 313.40 | 304.63 |

[a] Reference [36].

Table 2. Comparison of the lattice thermal conductivities of WTe$_2$ and WSe$_2$ obtained from various studies at 300 K.

| | WTe$_2$ (Wm$^{-1}$K$^{-1}$) | | WSe$_2$ (Wm$^{-1}$K$^{-1}$) | | |
|---|---|---|---|---|---|
| | This work | Experimental | This work | Theoretical | Experimental |
| [100] | 9.03 | | 15.00 | 50[b] | 9.7[c] |
| [010] | 7.69 | 0.8[a] | 14.59 | 50[b] | 9.7[c] |
| [001] | 0.45 | | 0.90 | 6[b] | 2.09[c] |

[a] Reference [24].

[b] Reference [16].

[c] Reference [43].

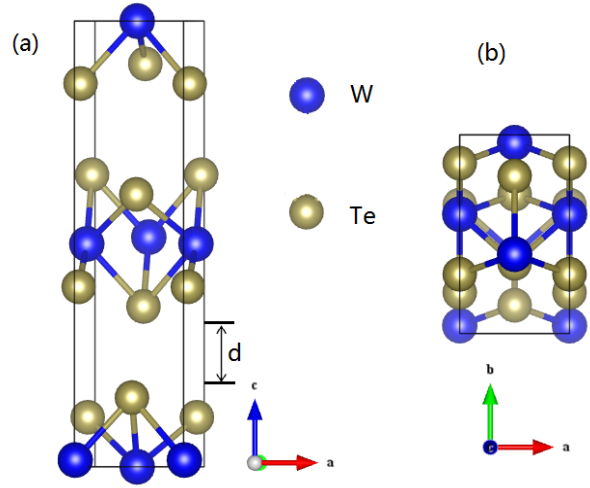

Figure 1. Crystal structure of $WTe_2$. The nearest interlayer distance ($d$) is shown. The side view and top view are shown in (a) and (b), respectively.

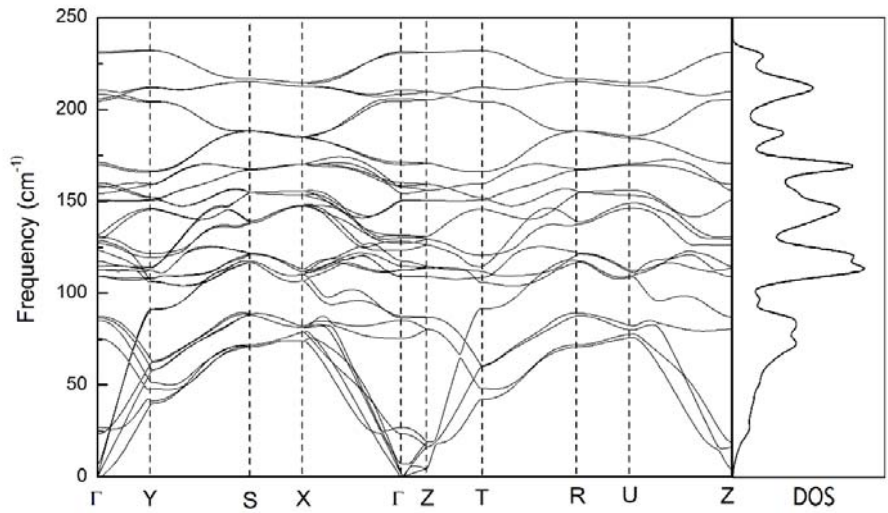

Figure 2. Calculated phonon dispersion and DOS for bulk $WTe_2$. The **k** points are $\Gamma = (000)$, $Y=(0\frac{1}{2}0)$, $S=(\frac{1}{2}\frac{1}{2}0)$, $X=(\frac{1}{2}00)$, $Z=(00\frac{1}{2})$, $T=(0\frac{1}{2}\frac{1}{2})$, $R=(\frac{1}{2}\frac{1}{2}\frac{1}{2})$, $U=(\frac{1}{2}0\frac{1}{2})$.

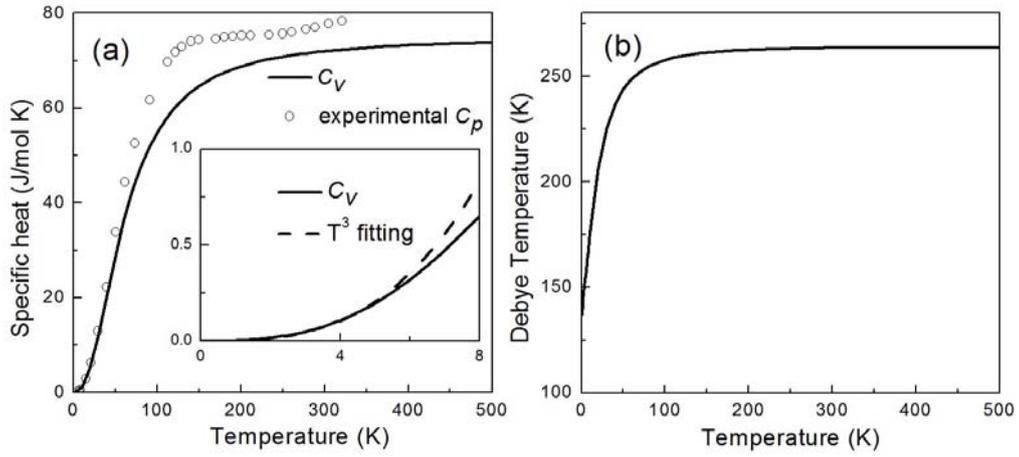

Figure 3. Temperature dependence of specific heat and Debye temperature. In (a) the solid line represents the constant–volume specific heat calculated. And the hollow circles mean the experimental constant–pressure specific heat from Ref. 22. The inset of (a) shows $C_v$ comparing with the $T^3$ fitting at low temperature. Debye temperature shows in (b).

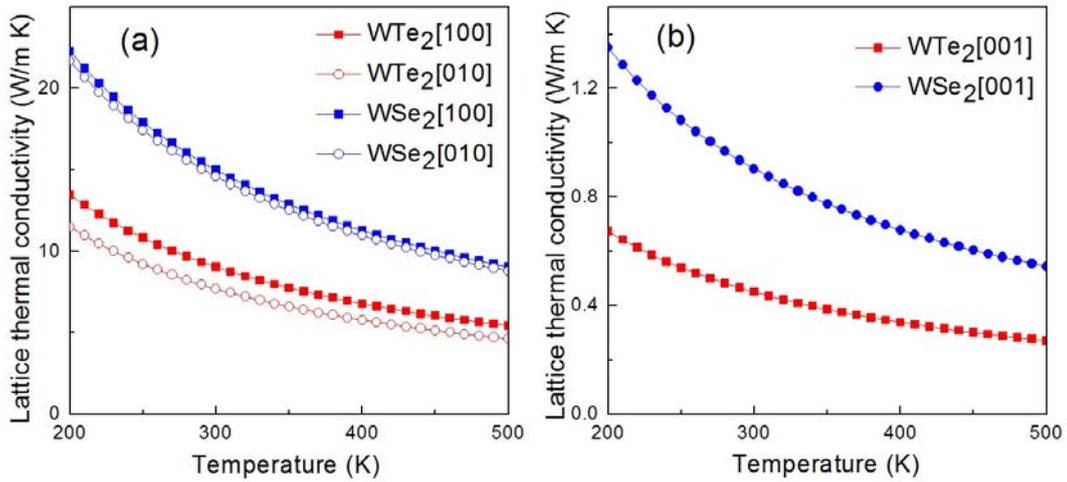

Figure 4. The temperature dependence of lattice thermal conductivity of $WTe_2$ and $WSe_2$. The thermal conductivities along [100] and [010] directions are shown in (a), and the contrast of thermal conductivities along [001] direction are shown in (b) individually.

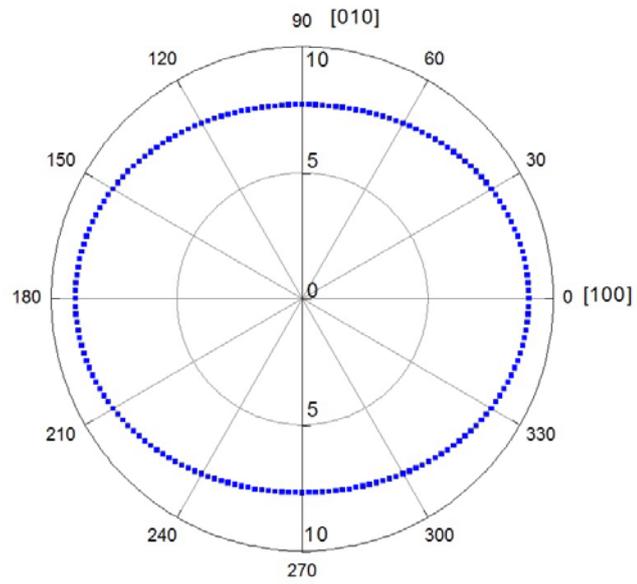

Figure 5. The lattice thermal conductivity $\kappa$ (W m$^{-1}$ K$^{-1}$) on （001）plane of WTe$_2$ at 300K.

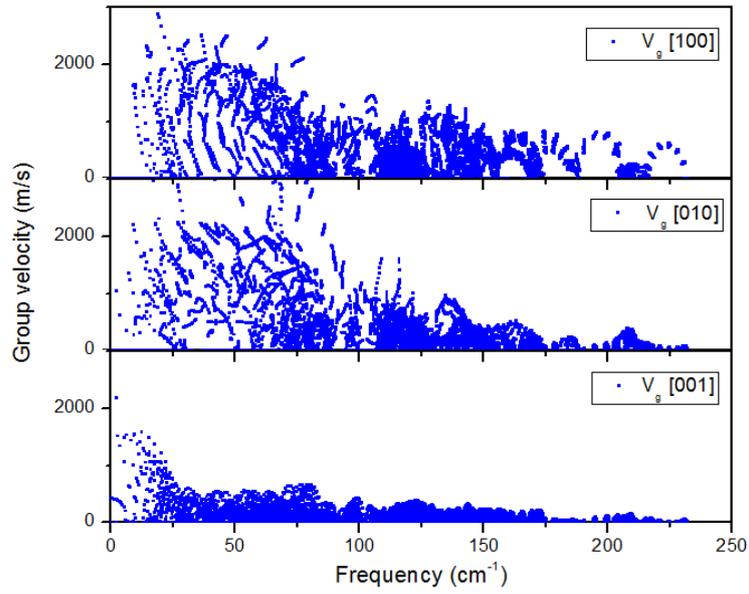

Figure 6. Group velocities of phonon along [100], [010], and [001] directions in WTe$_2$, respectively.

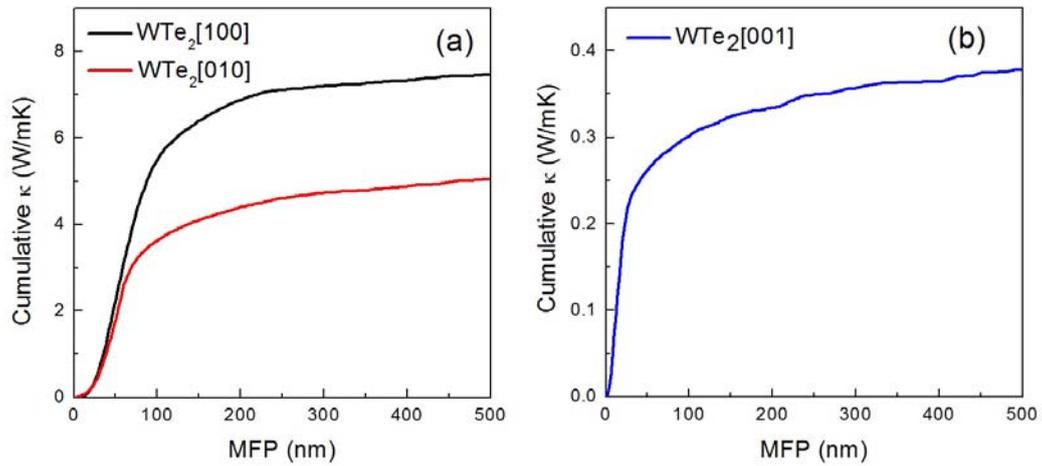

Figure 7. Cumulative thermal conductivity $\kappa$ with respect to phonon MFP in WTe$_2$ along [100], [010] and [001] directions, respectively.